\documentclass[%
 aip,
 amsmath,amssymb,
 reprint,%
]{revtex4-1}

\usepackage{graphicx}
\usepackage{dcolumn}
\usepackage{bm}

\usepackage[utf8]{inputenc}
\usepackage[T1]{fontenc}
\usepackage{mathptmx}
\usepackage{etoolbox}
\usepackage{subcaption}
\usepackage[colorlinks=true,urlcolor=blue,linkcolor=blue,citecolor=blue]{hyperref}
\usepackage{footmisc}
\usepackage[flushleft]{threeparttable}

\makeatletter
\def\@email#1#2{%
 \endgroup
 \patchcmd{\titleblock@produce}
  {\frontmatter@RRAPformat}
  {\frontmatter@RRAPformat{\produce@RRAP{*#1\href{mailto:#2}{#2}}}\frontmatter@RRAPformat}
  {}{}
}%
\makeatother
\begin{document}

\preprint{AIP/123-QED}

\title{Low-energy elastic scattering of electrons from 2H-pyran and 4H-pyran \\ with time delay analysis of resonances }
\author{Snigdha Sharma and Dhanoj Gupta$^{*}$\\Department of Physics, School of Advanced Sciences, Vellore Institute of Technology,\\ Vellore, Tamil Nadu, India - 632014\\}
 \email{dhanojsanjay@gmail.com}


\date{\today}

\begin{abstract}
Elucidating the significance of low-energy electrons in the rupture of DNA/RNA and the process involved in it is crucial in the field of radiation therapy. Capturing of the incident electron in one of the empty molecular orbitals and the formation of a temporary negative ion (TNI) is considered to be a stepping stone towards the lesion of DNA/RNA. This TNI formation manifests itself as a resonance peak in the cross-sections determined for the electron-molecule interaction. In the present work, we have reported the integral (ICS), differential (DCS), and momentum transfer (MTCS) cross-sections for the elastic scattering of low-energy electrons from the isomers, 2H-pyran and 4H-pyran $\rm{(C_5H_6O)}$, which are analogues of the sugar backbone of DNA. The single-center expansion method has been employed for the scattering calculations. Further, we have used the time delay approach to identify and analyze the resonance peaks. Our results for the ICS and DCS compare well with the only data available in the literature. MTCS data for 2H-pyran and 4H-pyran have been reported for the first time. Moreover, we have also identified an extra peak for each molecule, from time delay analysis, which might be a potential resonance.          
\end{abstract}

\maketitle


\section{\label{sec:level1}INTRODUCTION}

\begin{figure*}
\begin{subfigure}{0.45\textwidth}
\includegraphics[width=\textwidth]{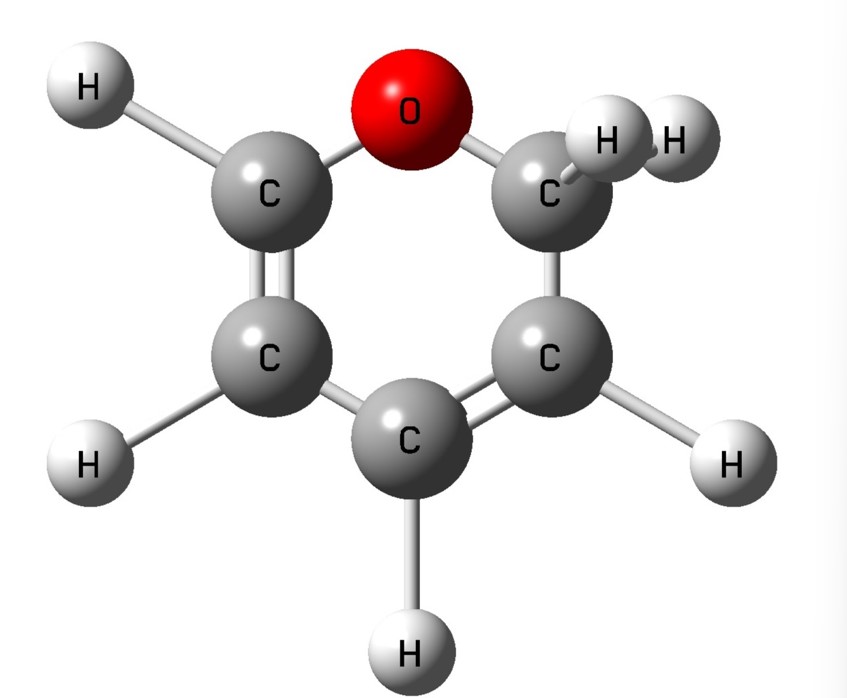}
\caption{}
\end{subfigure}
\begin{subfigure}{0.45\textwidth}
\includegraphics[width=\textwidth]{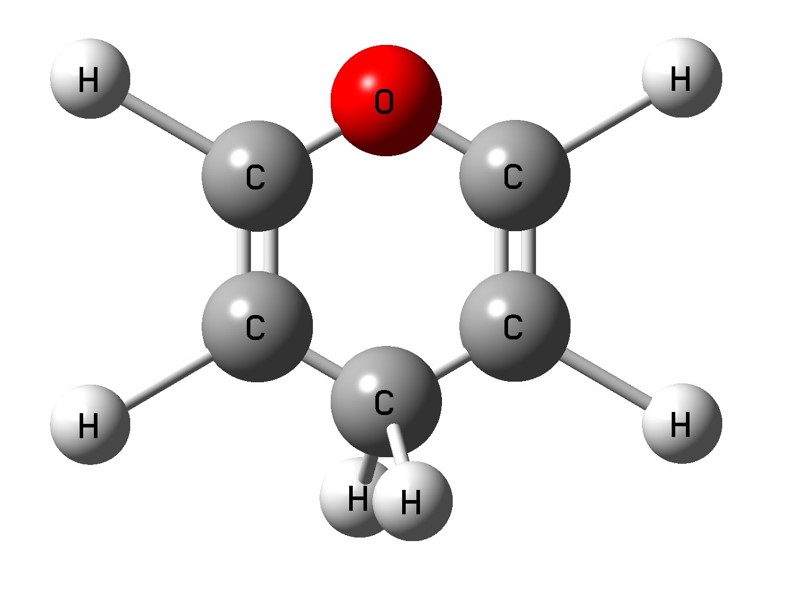}
\caption{}
\end{subfigure}
\caption{\label{fig:1} (a) 2H-pyran and (b) 4H-pyran molecular structure obtained using GaussView 6 graphical interface of Gaussian 16. \cite{g16}    }
\end{figure*}

Pyrans, the oxygen-containing six-membered ring compounds possessing two double bonds, are one of the most important structural subunits of natural products such as flavonoids in plants, macrocyclic molecules in aquatic life, secondary metabolites like xanthones, etc. \cite{ALMALKI2023998,YAVARI20032001} Pyran can be termed as 2H- and 4H-pyran depending upon the placement of the double bond in the ring structure.\cite{KUTHAN199519} In 2H-pyran the saturated carbon atom is in the second position whereas in 4H-pyran it is in the fourth position, as shown in Fig. \ref{fig:1}. 4H-pyran and its derivatives demonstrate an exemplary antibacterial 
\cite{KUMAR20093805}, antiviral 
\cite{SOLLIS19961805}, anticoagulant \cite{andreani1960aspects}, antimicrobial \cite{mahdavi2018synthesis}, antitumor \cite{mohr1975pyran}, and antiproliferative \cite{dell1998antiproliferative} properties. Pyran-based drugs are potent candidates in treating neurodegenerative diseases like Alzheimer’s.\cite{ALMALKI2023998} Analogues of 4h-pyran can bind efficiently to the minor groove of double-helical DNA assisting in the development of DNA-targeted drugs. \cite{RAMANA2016165} Steroidal 4H-pyrans have been developed as cancer chemotherapeutic agents.\cite{SHAMSUZZAMAN201336} 

Apart from the multiple therapeutic applications, 2H-pyran and 4H-pyran are considered to be the simple analogues of the sugar backbone of DNA and RNA.\cite{silva2024elastic} About two decades ago, Boudaïffa \textit{et al.} \cite{boudaiffa2000resonant} highlighted the significant role played by low-energy ($<$20 eV) electrons in the lesion of single- and double-strands of DNA/RNA. This instigated a series of work \cite{Bettega_furan,CHAUHAN2023110802,TAYEB2023117022,Sulzer,Singh} on the interaction of low-energy electrons with DNA/RNA components, especially their prototypes for experimental and theoretical vantage. The attractive polarization potential of the target molecule and the angular-momentum dependent repulsive potential give rise to a potential barrier that can capture the incoming electron. This resonance process forms a temporary negative ion which manifests itself as a peak in the cross-section calculation of the electron-molecule interaction. The captured electron may auto-detach itself from the molecule or transfer its energy leading to bond ruptures in different components of DNA/RNA, termed dissociative electron attachment.\cite{ingolfsson2019low} Aside from the lesion occurring due to the resonant capture of the low-energy electrons, electron impact ionization, for energies just above the ionization threshold, is also considered to be responsible for the fragmentation of biomolecules. \cite{S_Feil_2004,DENIFL200499} This has resulted in a series of works calculating the total and partial impact ionization cross-sections for molecules of biological relevance. \cite{dhanoj, suriya, vinodkumar}

In a recent work, Silva \textit{et al.} \cite{silva2024elastic} have investigated the resonant state formation of 2H-pyran and 4H-pyran by the interacting electrons employing the Schwinger multichannel method. \cite{SMC} In addition to the elastic scattering calculations, they have also reported the electronically inelastic and total ionization cross-sections of the two molecules. As per our knowledge, no other electron scattering calculation or experimental study is available in the literature for 2H-pyran and 4H-pyran. 
In order to locate the resonances, i.e., the formation of temporary negative anionic states during electron-molecule interaction, one can search for the resonant peaks in the calculated cross-section (elastic, inelastic, or total). However, the cross-section results might not give a clear picture of all the resonances present for the electron-molecule interaction due to the merging of closely placed peaks and masking of the physical resonances by non-resonant features. Time delay analysis, a method initially developed by Smith \cite{Smith}, is one of the most effective ways to clearly identify and characterize all the resonances. \cite{C7CP02916K}  In the context of quantum scattering, time delay corresponds to the excess electron density present in the vicinity of the molecule during interaction divided by the total incoming flux.\cite{Smith} 

In the present work, we have studied the low-energy electron interaction with 2H-pyran and 4H-pyran using the single-center expansion method. 
 \cite{gianturco1994generalized} We have also presented the time-delay analysis of the resonances present for the electron-molecule interaction. We have compared our calculated elastic integral (ICS) and differential (DCS) cross-section with the only available data in the literature of Silva \textit{et al}. \cite{silva2024elastic}. We have also compared the DCS with the experimental data available for benzene, which is a structurally similar molecule to the pyrans. Momentum transfer cross-section (MTCS) and time delay analysis are reported for the first time for the molecules.

\section{THEORETICAL BACKGROUD}
The theory employed to carry out the scattering calculations is based on the single-center expansion (SCE) approach. For a detailed explanation of the method, one can refer to the literature. \cite{BACCARELLI20111,Winifred} Here, we have highlighted only the important formulations of the SCE approach.  In this method, the bound and continuum wavefunctions, and potentials are expanded in terms of symmetry-adapted angular function about a single center, usually the center of mass of the target molecule. 
The one-electron bound wavefunction $\phi_i$ is expanded as, 

\begin{align}
 \mathcal \phi_{i}^{p\mu} (r)= \frac{1}{r} \left[\sum_{hl} u_{hl}^{i,p\mu} (r) X_{hl}^{p\mu} (\theta,\phi)\right].
\label{eqn1}
\end{align}
Here, index $i$ labels a specific multicentre bound orbital and $p$ is a particular irreducible representation (irrep) of the molecular point group. $\mu$ is one of the components of the irrep. The index $h$ denotes a basis function for a given partial wave $l$.  

A linear combination of the spherical harmonics results in the symmetry-adapted angular function,
\begin{align}
 X_{hl}^{p\mu} (\theta,\phi)=\sum_{m=-l}^{l} b_{hlm}^{p\mu}Y_{l}^{m} (\theta,\phi). 
\end{align}
The expansion coefficients $b$ can be determined using the character tables for each irrep of the molecular point group. 
A multicenter description of the molecular orbitals is required to expand Eq.\ref{eqn1}, which can be Gaussian-type orbitals (GTOs). The method of angular quadrature is then employed to get the radial coefficients subsequently,
 \begin{widetext}
\begin{align}
  u_{hl}^{i}(r;R)= \sum_{k,j,\nu,m} \int_{0}^{\pi}sin(\theta)d\theta \int_{0}^{2\pi} b_{hlm}^{i}Y_{l}^{m} (\theta,\phi)C_{kj}^{i}(R)d_{\nu}^{kj}g_{\nu}^{kj}(x_{k})d\phi,
\end{align}
 \end{widetext}
where $C_{kj}^{i}$ is the coefficient of $\mu^{th}$ GTO for $k^{th}$ atomic centre at a given molecular geometry R. $d_{\nu}^{kj}$ denotes the contraction coefficients of the cartesian gaussian functions $g_{\nu}^{kj}$. 

Once, the radial coefficients are known, the SCE one-electron bound wavefunction is used to calculate the electron density and hence the electrostatic potential, model exchange, and correlation-polarization potential. Finally, we get a set of coupled radial equations for the scattering electron,

\begin{align}
 \left[\frac{d^2}{dr^2}-\frac{l(l+1)}{r^2}\right] F_{lh}^{p\mu}(r)= 2\sum_{l'h'}V_{lh,l'h'}^{p\mu}(r)F_{l'h'}^{p\mu}(r),
 \label{eqn4}
\end{align}
which are solved to get the scattering parameters, viz, K- and T-matrix elements in the body-fixed frame. Here, $V_{lh,l'h'}^{p\mu}(r)$ is the coupling potential given as,

\begin{align}
 V_{lh,l'h'}^{p\mu}(r) = \int X_{hl}^{p\mu}(\theta,\phi)V(\textbf{r})X_{h'l'}^{p\mu}(\theta,\phi)d\theta d\phi  
\end{align}
\begin{align}
and, V(\textbf{r}) = V_{st} + V_{corr,pol} + V_{exc} \nonumber
\end{align}
The static potential ($V_{st}$) is calculated exactly using the molecular electron density $\rho(\textbf{r})$, whereas, a local model is considered for correlation-polarization ($V_{corr,pol}$)\cite{SCElib}, where, the short-range correlation potential, as given by Perdew and Zunger \cite{Perdew}, is used,

 \begin{widetext}
\begin{align}
  V_{corr} (\textbf{r}) & = lnr_s(0.0311+0.00133r_s) - 0.0084r_s - 0.0584 \qquad     \rm{for} \; r_s < 1.0 ,\\
                     & = \frac{\gamma(1+\frac{7}{6}\beta_1r_s^{1/2}+\frac{4}{3}\beta_2r_s)}{(1+\beta_1r_s^{1/2}+\beta_2r_s)^2} \qquad\rm{for} \;r_s \geq 1.0.
\end{align}
 \end{widetext}
Here, $\gamma$ = -0.1423, $\beta_1$ = 1.0529, $\beta_2$ = 0.334, and $r_s$ = $[\frac{3}{4\pi\rho(\textbf{r})}]^{1/3}$.
Moreover, the following long-range form of polarization potential, which depends upon the polarizability $\alpha$ of the target molecule, is used,

\begin{align}
  V_{pol} (\textbf{r})  = - \frac{1}{2r^6} \sum_{i,j}^3 q_iq_j\alpha_{ij},
\end{align}
where $q_i,q_j$ = x, y, z.
For the exchange potential ($V_{exc}$), the energy-dependent Hara type free electron gas model is employed.\cite{Hara,Volscat} 

For non-polar molecules, the rotationally elastic (rotationally resolved) DCS for no transition in the rotational states ($n=0\rightarrow n'=0$) and rotationally inelastic (rotationally unresolved) DCS for transition between different rotational states ($n\rightarrow n'$) in the space-fixed frame is computed as,

\begin{align}
 \frac{d\sigma}{d\Omega}(n\rightarrow n';\theta)= \sum_{L=0}^{L=LBIG}A_L(n\rightarrow n')P_L(cos\theta),
 \label{eqn6}
\end{align}
where $A_L$ is the expansion coefficient which depends upon the energy of the incoming projectile and is determined from the K-matrix elements, $LBIG$ is the maximum number of partial waves considered for the continuum electron, and $P_L(cos\theta)$ is the Legendre polynomial for a scattering angle $\theta$. The overall sum of the rotationally resolved and unresolved DCS gives the rotationally summed DCS.

In the case of polar molecules, the presence of a long-range dipole interaction between the projectile electron and the target results in a slow convergence of the sum in Eq.\ref{eqn6}. To overcome this problem the following closure formula is used, 

\begin{align}
 \frac{d\sigma}{d\Omega}(n\rightarrow n';\theta)= \frac{d\sigma^B}{d\Omega} +  \sum_{L=0}^{L=LBIG}(A_L - A_L^B)P_L(cos\theta),
 \label{eqn7}
\end{align}
where the ones with a superscript $B$ are calculated using the first Born approximation resulting in a born corrected DCS for polar molecules (Eq.\ref{eqn7}). The corresponding elastic ICS is computed using,

\begin{align}
 \sigma = \sigma_{rd}^B + \sigma_{cc} - \sigma_{fd}^B
 \label{eqn8}
\end{align}
Here, $\sigma_{rd}^B$ is the ICS calculated for a rotating dipole within the first Born approximation, $\sigma_{cc}$ is the result of close-coupling calculation under fixed-nuclei approximation, and $\sigma_{fd}^B$ is the ICS for a fixed dipole.

Similarly, the MTCS is calculated using,

\begin{align}
 \sigma^M= \sigma_{rd}^{M(B)} + \sigma_{cc}^M - \sigma_{fd}^{M(B)}
 \label{eqn9}
\end{align}

The general formula used for calculating the K-matrix elements, DCS, ICS, and MTCS under first Born approximation, as reported by Sanna and Gianturco \cite{Polydcs}, is only valid for molecules possessing higher symmetry (e.g. $\mathrm{C_{2v}}$ ) and dipole moment aligned along the symmetry axis. For lower point group molecules (e.g. $\mathrm{C_{1}}$ or $\mathrm{C_{s}}$), the formula needs to be modified as suggested in the work reported by Franz \textit{et al.} \cite{franz2014low} Below are the modified equations used to perform Born correction for low-symmetry polar molecules. 

The Born K-matrix elements are calculated using,

\begin{align}
    {K}^{p\mu}_{lh,l^\prime h^\prime} = -2k \int_{0}^{\infty} j_l(kr)j_{l^\prime}(kr)r^2dr \langle X_{lh}^{p\mu}| \sum_{\lambda H}v_{\lambda H}(\textbf{r})X_{\lambda H}^A|X_{l^\prime h^\prime}^{p\mu} \rangle ,
\label{eqn10}
\end{align}
where for low-symmetry molecules, the potential term includes all three components of the dipole moment, i.e,

\begin{align}
\sum_{\lambda H}v_{\lambda H}(\textbf{r})X_{\lambda H}^A = \frac{A_{-1}}{r^2}S_1^{-11} + \frac{A_{0}}{r^2}S_1^{01} + \frac{A_{1}}{r^2}S_1^{11}.
 \label{eqn11}
\end{align}
Here, $S_l^{mq}$ are the real spherical harmonic functions with $l,m$ being the angular momentum quantum numbers. The coefficients $A_m$ explicitly depend upon the components of dipole moment, $D_m$, as,

\begin{align} 
A_m = - \left( \frac{4\pi}{3} \right)^\frac{1}{2} D_m    \nonumber.
\end{align}

Further, the DCS calculated under Born-approximation for low-symmetry molecules is given as,

\begin{align} 
\frac{d\sigma}{d\Omega}(n:0\rightarrow1) = \sum_{\tau^\prime}  \frac{4D_{\tau^\prime}^2}{3} \frac{k_{1\tau^\prime}}{k_{00}} \frac{1}{\left( k_{00}^2 + k_{1\tau^\prime}^2 - 2k_{00}k_{1\tau^\prime}\cos\theta\right)} , 
 \label{eqn12}
\end{align}
where the summation is over the quantum number $\tau^\prime$ which represents the sublevels of the rotational level $n=1$. $D_{\tau^\prime}$ is the component of the dipole moment corresponding to the quantum number  $\tau^\prime$. $\theta$ is the angle between the initial and final momenta of the projectile electron, where, these momenta are given by,

\begin{align} 
k_{00} = \sqrt{2E} \nonumber, \qquad k_{1\tau^\prime} = \sqrt{2[E-(\epsilon_{1\tau^\prime} - \epsilon_{00})]} \nonumber.
\end{align}
Here, $E$ is the energy of the incident electron, $\epsilon_{00}$ is the energy of the rotational ground state and $\epsilon_{1\tau^\prime}$ is the energy of the rotational level $n=1, \tau^\prime=-1,0,1$. Only the rotational transition $n:0 \rightarrow 1$ is considered for calculating Born DCS because the contribution from all other transitions involving only the dipole term is zero. \cite{Polydcs}

In a similar way, the Born ICS and MTCS for low-symmetry molecules take the form,

\begin{align} 
\sigma(n:0\rightarrow1) = \frac{8\pi}{3k_{00}^2}\sum_{\tau^\prime} D_{\tau^\prime}^2 \rm{ln} \left|\frac{k_{00} + k_{1\tau^\prime}}{k_{00} - k_{1\tau^\prime}} \right| 
 \label{eqn13}
\end{align}
and

\begin{align} 
\sigma^M(n:0\rightarrow1) = \frac{8\pi}{3k_{00}^2}\sum_{\tau^\prime} D_{\tau^\prime}^2 \left( 1 - \frac{(k_{00} - k_{1\tau^\prime})^2}{2k_{00}k_{1\tau^\prime}} \rm{ln} \left|\frac{k_{00} + k_{1\tau^\prime}}{k_{00} - k_{1\tau^\prime}} \right| \right),
 \label{eqn14}
\end{align}
respectively, with the terms having the usual meaning as given in Eq.\ref{eqn12}. 

For the analysis of the resonances, the time-delay matrix or the Q-matrix is first constructed from the scattering matrix or the S-matrix as follows,

\begin{align}
 Q(E)=-i\hbar\frac{dS}{dE}S^\dagger,
 \label{eqn15}
\end{align}
where the dagger denotes the Hermitian conjugate of the S-matrix and $\rm{-i\frac{d}{dE}}$ is the time-operator. The eigenvalues of the Q-matrix correspond to the time delay experienced by the interacting electrons. The maximum eigenvalue denotes the largest time delay and hence is sensitive to depicting the resonances. The maximum eigenvalue is plotted against the energy of the incoming electron, where, the resonances can be identified as Lorentzian peaks. 

\begin{table*}
\caption{\label{tab:table1}Comparison of the SCE computed and the theoretical dipole moment of targets (Units in Debye). }
\begin{ruledtabular}
\begin{tabular}{ccccc}
Target&Point group&Computed number of electrons (SCE)&Computed dipole moment (SCE) 
&Compared dipole moment\\ \hline
 2H-pyran\\ $(C_5H_6O)$&$C_{s}$&44.005 &0.90946 & 0.89810\footnote{\label{a}Gaussian 16\cite{g16}}, 0.94\footnote{\label{b}Silva \textit{et al.}\cite{silva2024elastic}} \\
4H-pyran\\ $(C_5H_6O)$&$C_{2v}$ & 44.011 & 0.92017 & 0.91500\footref{a}, 1.14\footref{b}\\

\end{tabular}
\end{ruledtabular}
\end{table*}

\begin{table*}
\caption{\label{tab:table1a}Molecular properties of the targets used in present calculations. Dipole moments and Rotational constants are calculated using the Hartree-Fock level of theory with 6-31G(d,p) basis set in Gaussian 16 software. \cite{g16} Rotational energy levels are generated using ASYMTOP code. \cite{JAIN1983301}  Ionization potential is taken from literature. \cite{silva2024elastic} Polarizability for both 2H-pyran and 4H-pyran is imported from the ChemSpider website. \cite{CHEMSPIDER} }
\begin{ruledtabular}
\begin{tabular}{ccccc}
Target & 2H-pyran $(C_5H_6O)$ & 4H-pyran $(C_5H_6O)$\\ \hline
Dipole moment in Debye \\ \hline
$\mu_x$ & 0.3379  & 0.0  \\
$\mu_y$ & -0.8321  & 0.0  \\
$\mu_z$ & 0.0  & -0.9150 \\
$|\mu|$ & 0.8981 & 0.9150  \\
\hline
Rotational constant in GHz \\ \hline
$A$ & 5.55903 & 5.77396  \\
$B$ & 5.42707 & 5.25989 \\
$C$ & 2.79288 & 2.79884 \\
\hline
Rotational energy levels in $10^{-6}$ eV \\ \hline
$\epsilon_{1-1}$ & 33.9950 & 33.3280 \\
$\epsilon_{10}$ & 34.5400 & 35.4540  \\
$\epsilon_{11}$ & 45.4350 & 45.6320 \\
\hline
Ionization potential in eV \\ \hline
I.P. & 7.85 & 8.63 \\
\hline
Polarizability in a.u.\\ \hline
$\alpha$ & 64.17 & 64.17 \\
\end{tabular}
\end{ruledtabular}
\end{table*}

\section{COMPUTATIONAL DETAILS}

Multicentre GTOs and optimized geometry for the target molecules are generated using the quantum chemistry software Gaussian 16. \cite{g16} All the calculations in the Gaussian are performed within the Hartree-Fock level of theory using the 6-31G(d,p) basis set. The symmetries for 2H-pyran and 4H-pyran are fixed at their natural point groups of $\mathrm{C_{s}}$ and $\mathrm{C_{2v}}$, respectively. Subsequently, the single-center expanded bound state wavefunctions, electron density, and potentials are generated using SCELib 4.0 \cite{SCElib}, which is a set of codes developed by Sanna \textit{et al.} For both 2H- and 4H-pyran, the maximum number of the partial waves, $l_{max}$, given in Eq.\ref{eqn1}, is set at 50 to ensure a satisfactory convergence of the molecular properties with all the molecular orbitals normalized approximately to unity. The radial and angular grids chosen for integration correctly calculated the number of electrons in the molecules and gave a satisfactory value of their dipole moments. The theoretical value of polarizability, $\alpha$ = 64.17 $\rm{a_0^3}$, extracted from the ChemSpider website \cite{CHEMSPIDER} which uses ACD/Labs Percepta Platform to calculate the physicochemical properties\cite{ACDlab}, is used for both 2H-pyran and 4H-pyran as the experimental value for the same is unavailable. Polarizability is essential for correctly modeling the polarization potential in the asymptotic regions. The ionization potential values considered are 0.2885 a.u. and 0.31714 a.u, as reported by Silva \textit{et al.} \cite{silva2024elastic}, for the calculation of the exchange potential for 2H-pyran and 4H-pyran, respectively.  As the experimental dipole moment for the molecules is unavailable, we, therefore, compared the calculated dipole moment using SCE approach with the one computed using the Hartree-Fock level of theory with 6-31G(d,p) basis set in Gaussian 16 software \cite{g16} and the dipole moment values reported by Silva \textit{et al.} \cite{silva2024elastic} which are calculated using the
second-order Møller–Plesset perturbation theory (MP2) and the
aug-cc-pVDZ basis set. A good comparison of dipole moments, refer to Table \ref{tab:table1}, for both the molecules hints towards the satisfactory modeling of the target molecules.



The coupled radial equations for scattering (Eq.\ref{eqn4}) are then solved using VOLSCAT 2.0 \cite{Volscat}, which is another set of codes, giving ICS as the output. A good number of partial waves are considered for the scattering calculation in order to achieve convergence in the ICS result. The time-delay matrix is also constructed using VOLSCAT 2.0. However, the determined ICS is in the body-fixed frame. Another code set, the POLYDCS \cite{Polydcs}, is used to finally obtain the rotationally summed DCS and ICS in the space-fixed frame. For the frame transformation, the ASYMTOP \cite{JAIN1983301}  code is used to generate the molecules' rotational energy levels and eigenfunctions. The first six rotational transitions ( $\rm{0\rightarrow0, 0\rightarrow1, 0\rightarrow2, 0\rightarrow3, 0\rightarrow4, and \: 0\rightarrow5}$) are considered for both molecules. The molecular properties used for performing different calculations are summarized in Table \ref{tab:table1a}.  

Additionally, the original POLYDCS code suite \cite{Polydcs} has been developed to perform calculations for molecules having either an axis of symmetry with the dipole moment aligned along it or a zero dipole moment. For lower point group ($\mathrm{C_{s}}$ or $\mathrm{C_{1}}$) polar molecules, like 2H-pyran in this case, we have modified the code to include the relevant equations (refer to Eqs.\ref{eqn11}--\ref{eqn14}) for the first Born approximation calculations as suggested by Franz \textit{et al.} in their work. \cite{franz2014low} To confirm the correct implementation of the equations in the code, we successfully reproduced the integral cross-section for elastic scattering of positron by Thymine nucleobase reported by Franz \textit{et al.} \cite{franz2014low} The reproduced data is not presented in this work.   

\begin{figure}
\includegraphics[width=0.55\textwidth,height=1.5\columnwidth]{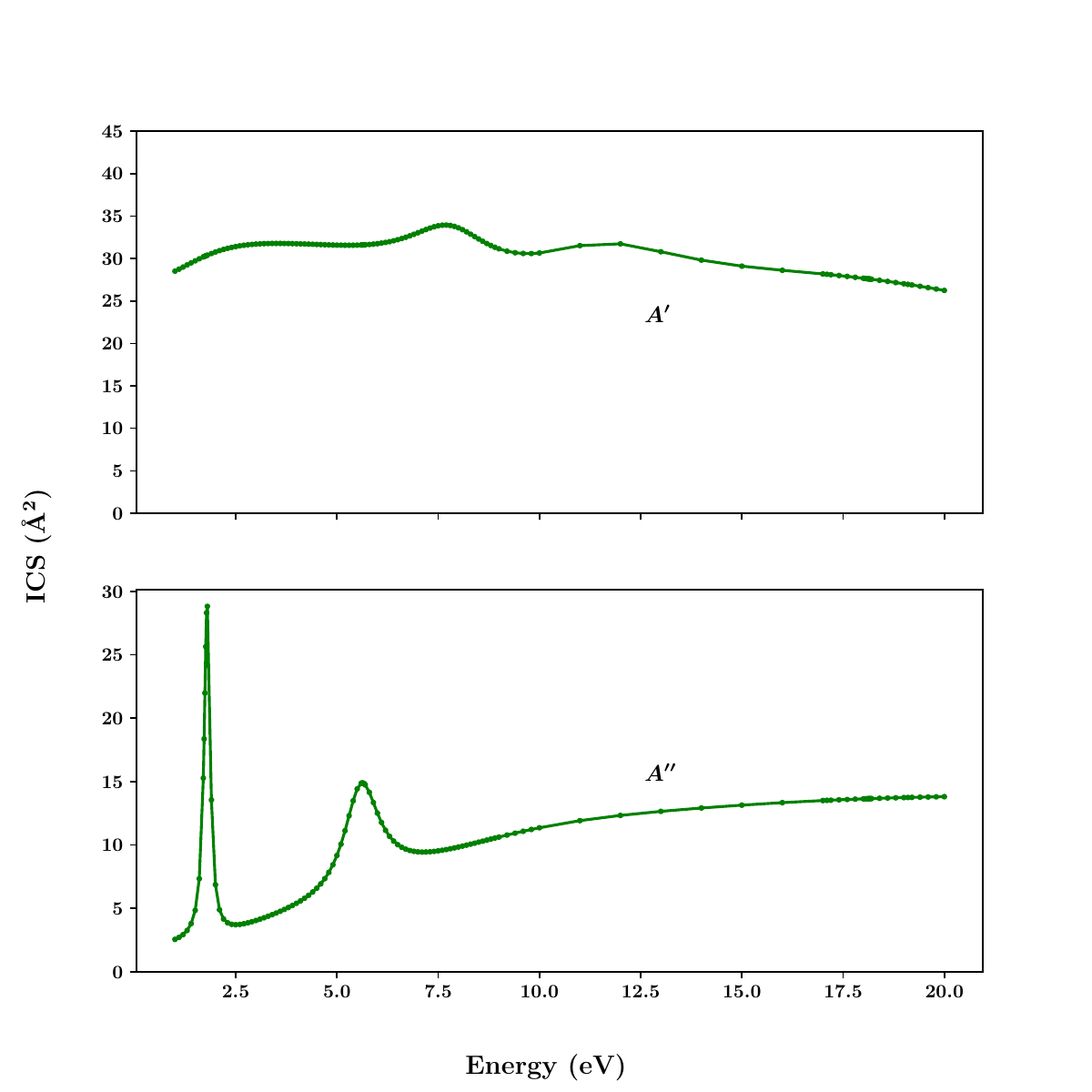}
\caption{\label{fig:2} Irrep-wise contribution to integral elastic cross-section of 2H-pyran.}
\end{figure}

\begin{figure}
\includegraphics[width=0.55\textwidth,height=1.5\columnwidth]{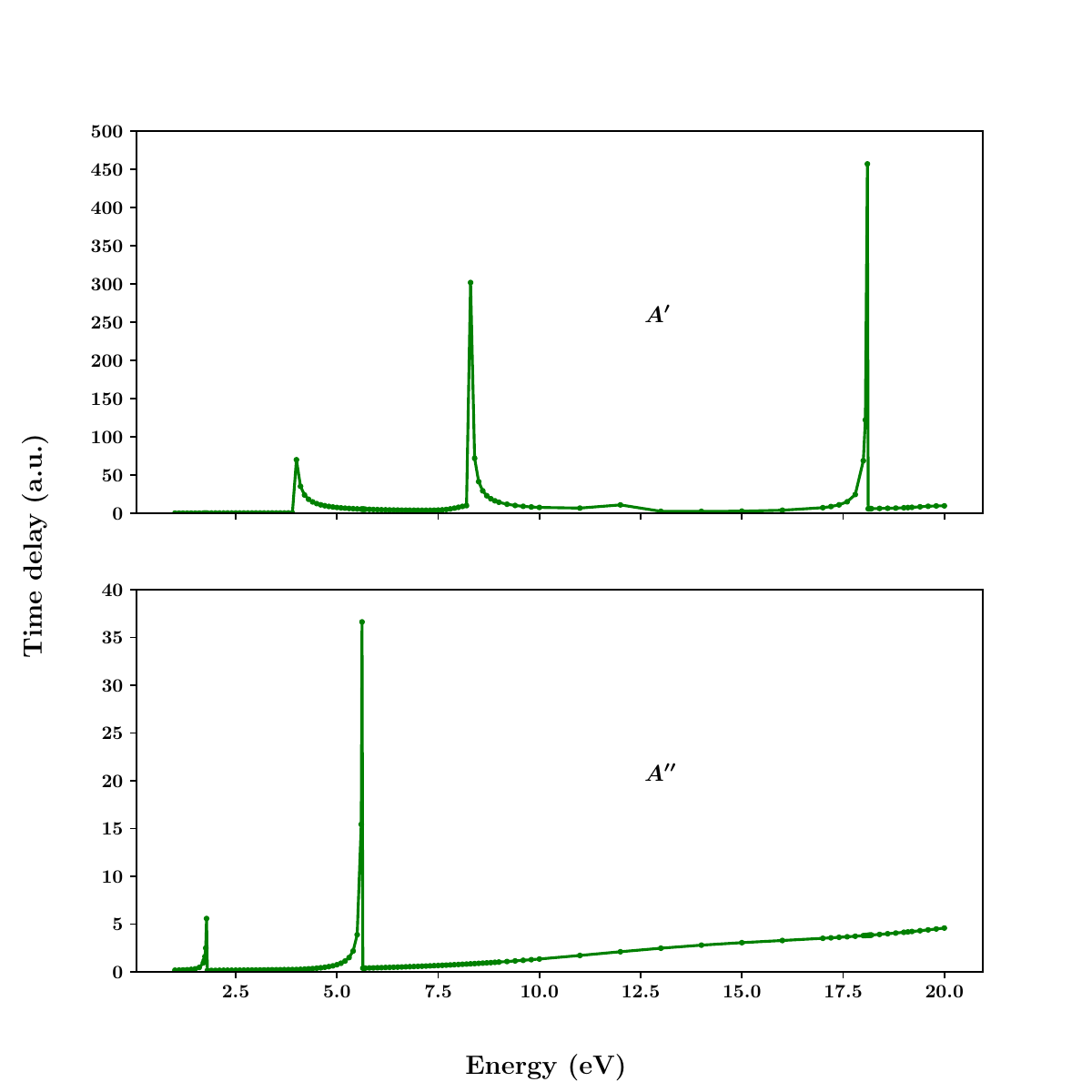}
\caption{\label{fig:3} Irrep-wise maximum eigenvalue of the time delay matrix of 2H-pyran.}
\end{figure}

\begin{figure*}
\begin{subfigure}{0.45\textwidth}
\includegraphics[width=\textwidth]{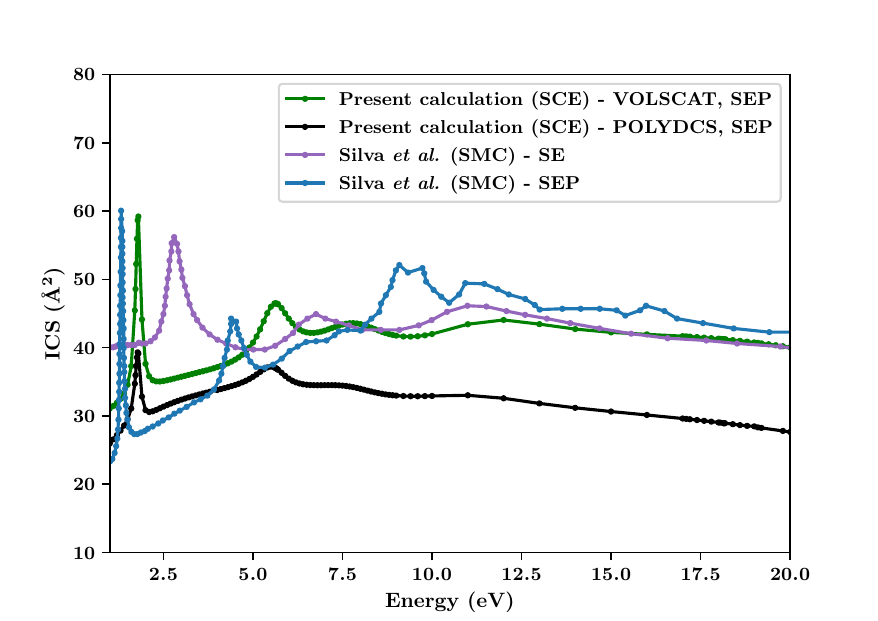}
\caption{}
\end{subfigure}
\begin{subfigure}{0.45\textwidth}
\includegraphics[width=\textwidth]{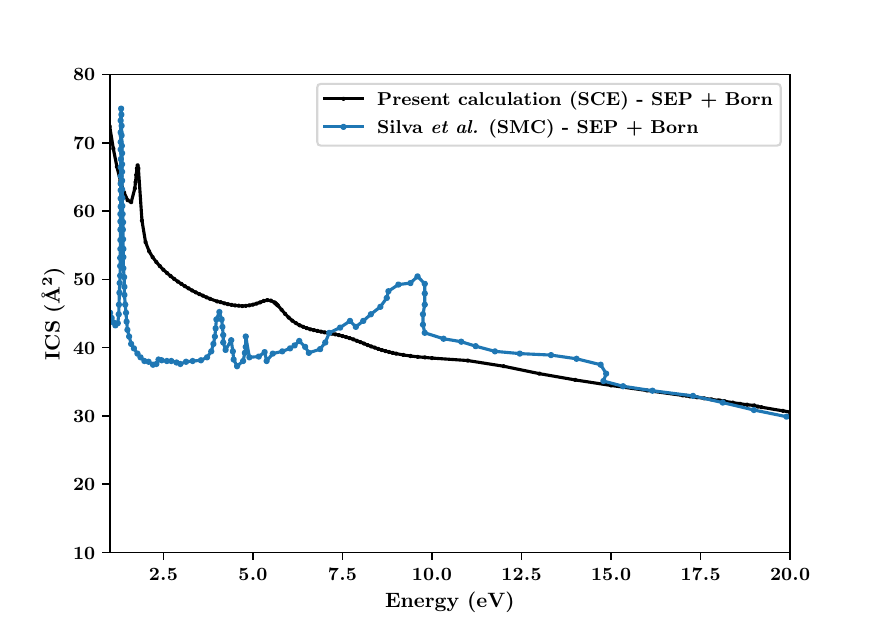}
\caption{}
\end{subfigure}
\caption{\label{fig:4} (a) Born-uncorrected integral elastic cross-section for the scattering of electrons by 2H-pyran. Green dotted-line: present result in body-fixed frame (VOLSCAT calculation). Black dotted-line: present result in space-fixed frame (POLYDCS calculation). Purple dotted-line: SMC calculation of Silva \textit{et al.} \cite{silva2024elastic} under 1-channel SE approximation. Blue dotted-line: SMC calculation of Silva \textit{et al.} \cite{silva2024elastic} under 1-channel SEP approximation. (b) Born-corrected integral elastic cross-section for the scattering of electrons by 2H-pyran. Black dotted-line: present result in space-fixed frame (POLYDCS calculation). Blue dotted-line: SMC calculation of Silva \textit{et al.} \cite{silva2024elastic} under up to 129-channel SEP approximation.  }
\end{figure*}

\begin{figure}
\includegraphics[width=0.55\textwidth,height=.4\textwidth]{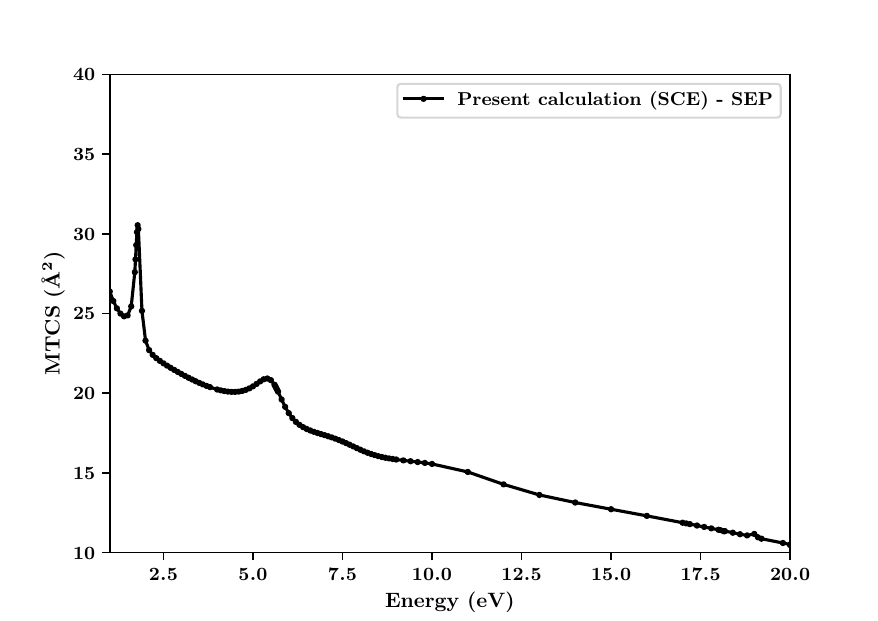}
\caption{\label{fig:5} Born-corrected momentum transfer cross-section for the scattering of electrons by 2H-pyran}
\end{figure}

\begin{figure*} 
\begin{subfigure}{0.45\textwidth}
\includegraphics[width=\textwidth]{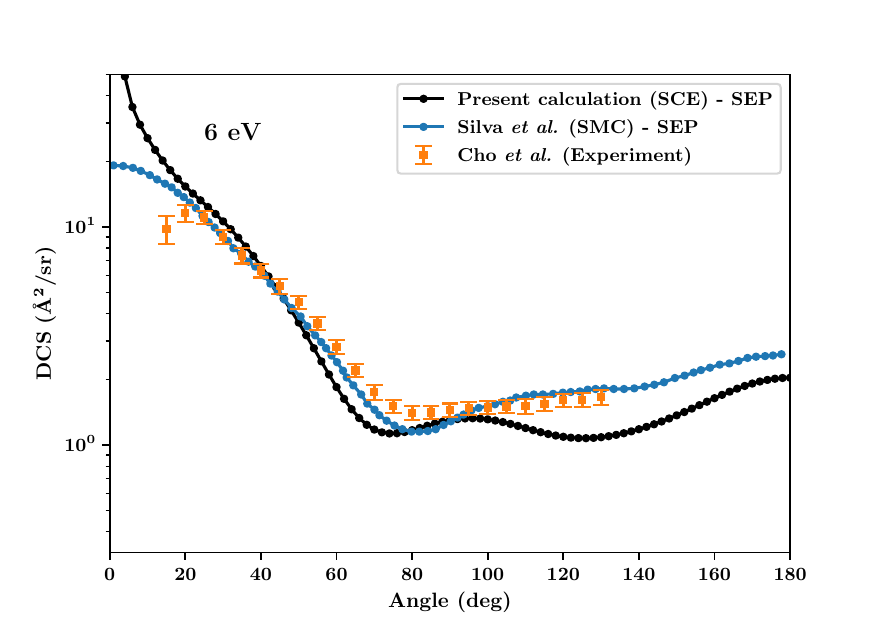}
\caption{}
\end{subfigure}
\begin{subfigure}{0.45\textwidth}
\includegraphics[width=\textwidth]{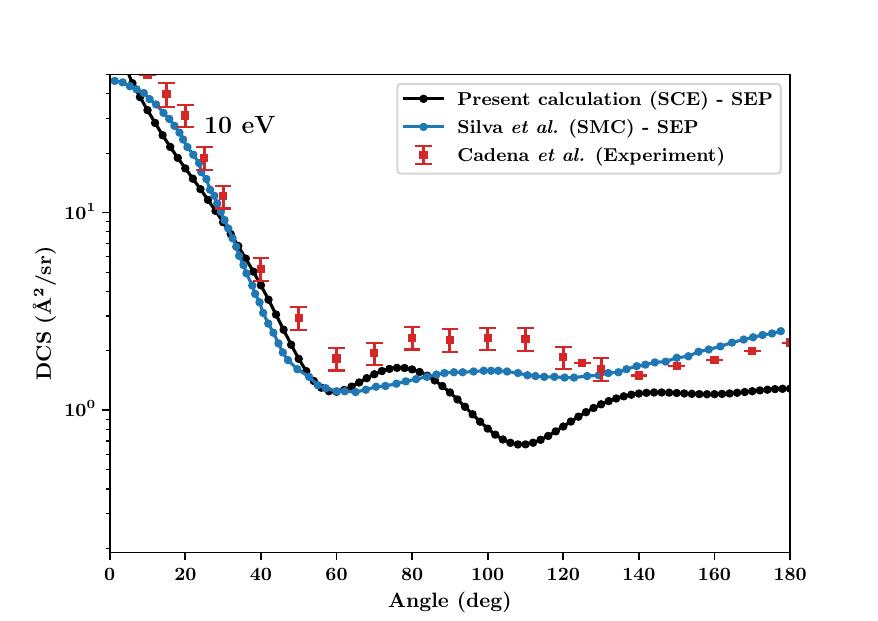}
\caption{}
\end{subfigure}
\begin{subfigure}{0.45\textwidth}
\includegraphics[width=\textwidth]{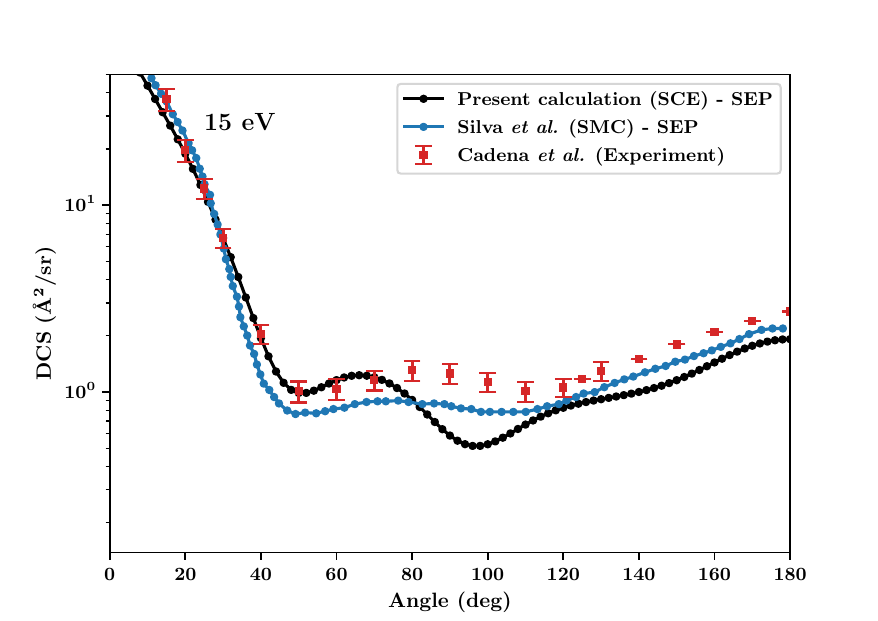}
\caption{}
\end{subfigure}
\caption{\label{fig:6} Born-corrected differential cross-section for the scattering of electrons by 2H-pyran at the incident energies of (a) 6 eV (b) 10 eV (c) 15 eV. Black dotted-line: present work. Blue dotted-line: SMC calculation of Silva \textit{et al.} \cite{silva2024elastic} under 37-channel SEP approximation for 6 eV and 10 eV DCS, and 108-channel SEP approximation for 15 eV DCS. Red squares with error: experimental DCS measured by Cadena \textit{et al.}\cite{cadena2022cross} for benzene molecule. Orange squares with error: experimental DCS measured by Cho \textit{et al.}\cite{Cho_2001} for benzene molecule. }
\end{figure*}

\begin{figure}
\includegraphics[width=0.55\textwidth,height=1.5\columnwidth]{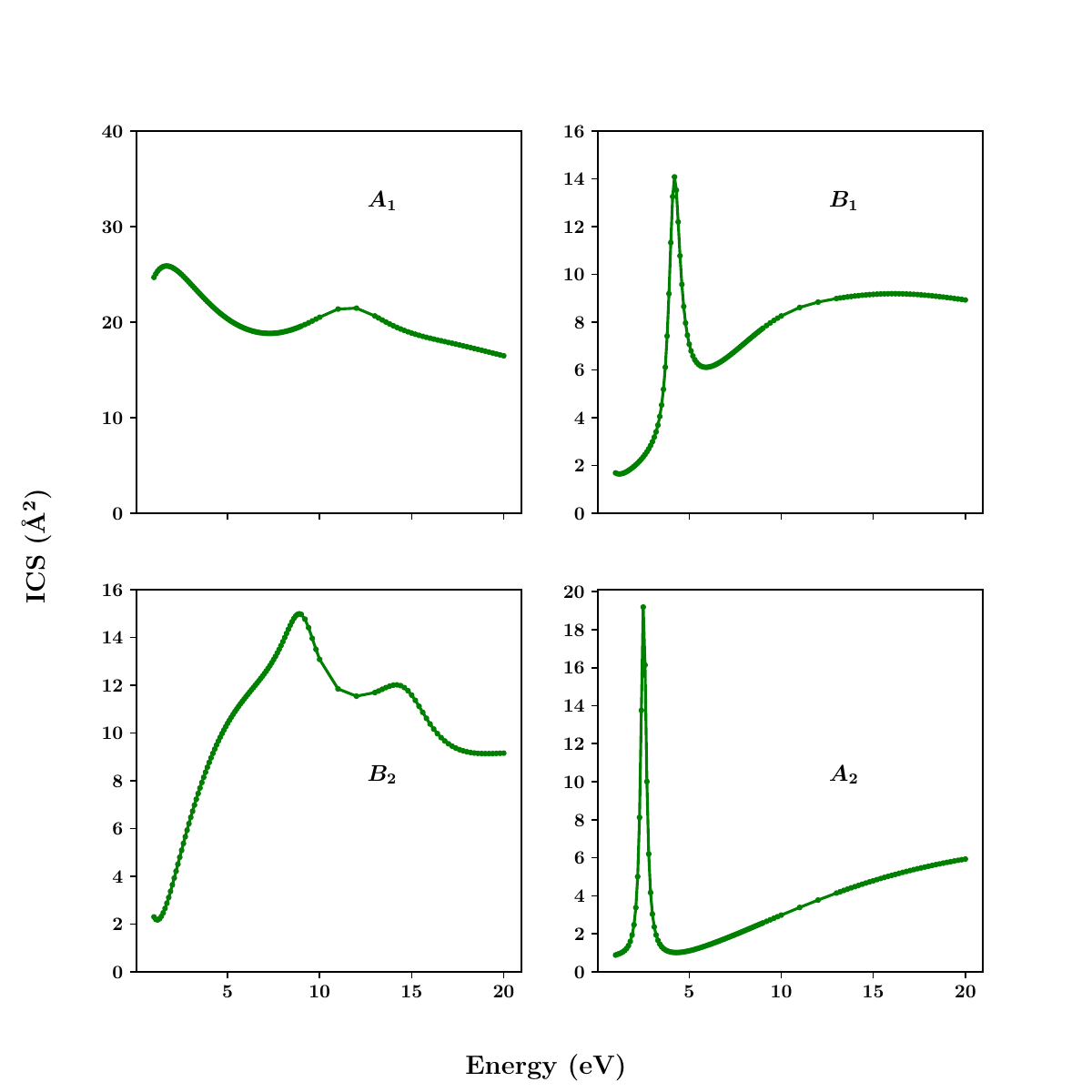}
\caption{\label{fig:7} Irrep-wise contribution to integral elastic cross-section of 4H-pyran.}
\end{figure}

\begin{figure}
\includegraphics[width=0.55\textwidth,height=1.5\columnwidth]{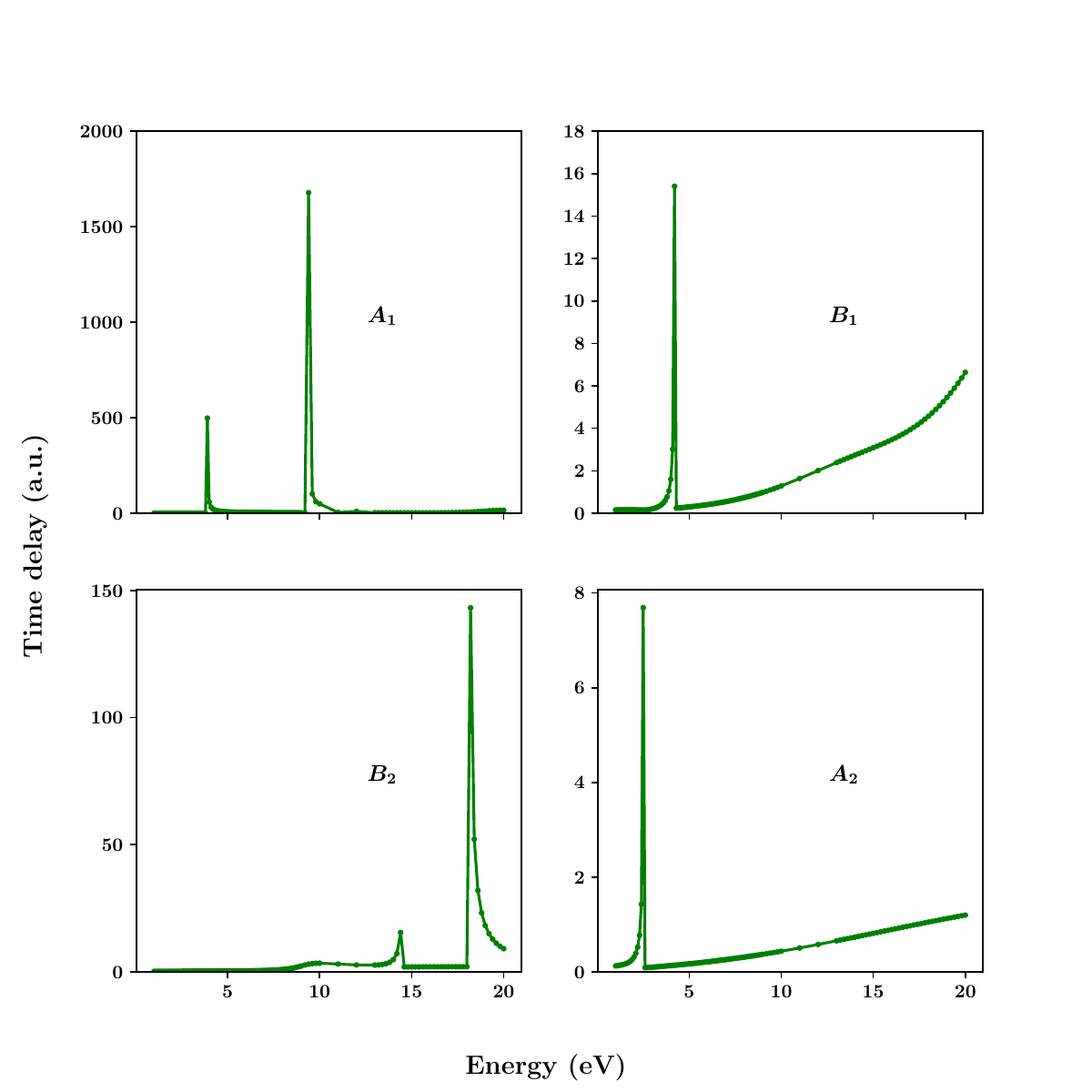}
\caption{\label{fig:8} Irrep-wise maximum eigenvalue of the time delay matrix of 4H-pyran.}
\end{figure}

\begin{figure*}
\begin{subfigure}{0.45\textwidth}
\includegraphics[width=\textwidth]{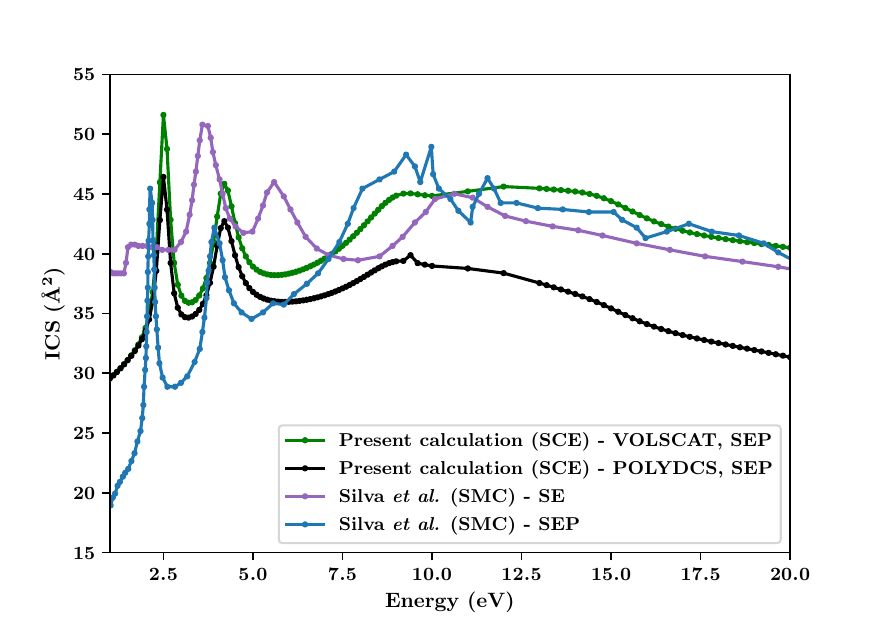}
\caption{}
\end{subfigure}
\begin{subfigure}{0.45\textwidth}
\includegraphics[width=\textwidth]{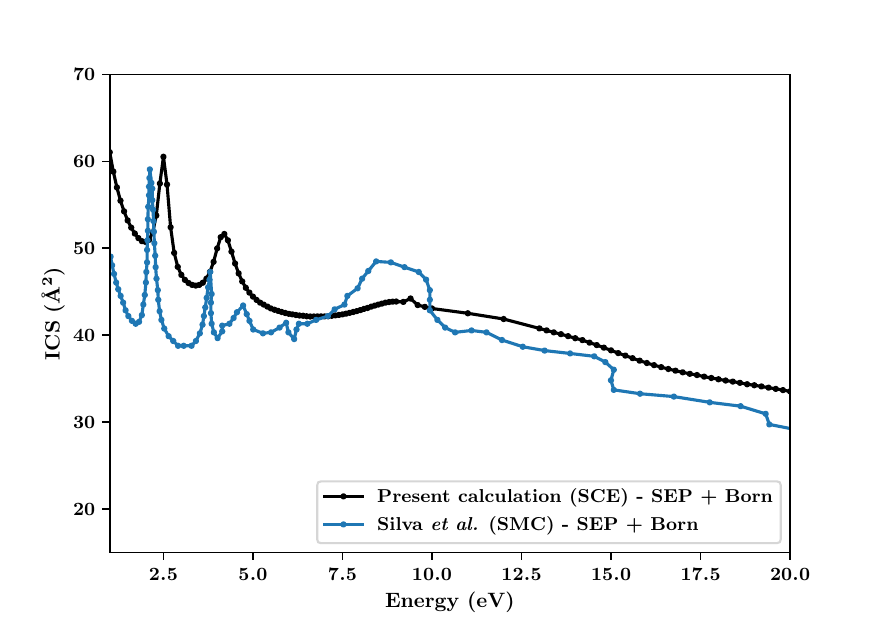}
\caption{}
\end{subfigure}
\caption{\label{fig:9} (a) Born-uncorrected integral elastic cross-section for the scattering of electrons by 4H-pyran. Green dotted-line: present result in body-fixed frame (VOLSCAT calculation). Black dotted-line: present result in space-fixed frame (POLYDCS calculation). Purple dotted-line: SMC calculation of Silva \textit{et al.} \cite{silva2024elastic} under 1-channel SE approximation. Blue dotted-line: SMC calculation of Silva \textit{et al.} \cite{silva2024elastic} under 1-channel SEP approximation. (b) Born-corrected integral elastic cross-section for the scattering of electrons by 4H-pyran. Black dotted-line: present result in space-fixed frame (POLYDCS calculation). Blue dotted-line: SMC calculation of Silva \textit{et al.} \cite{silva2024elastic} under up to 249-channel SEP approximation.  } 
\end{figure*}

\begin{figure}
\includegraphics[width=0.55\textwidth,height=.4\textwidth]{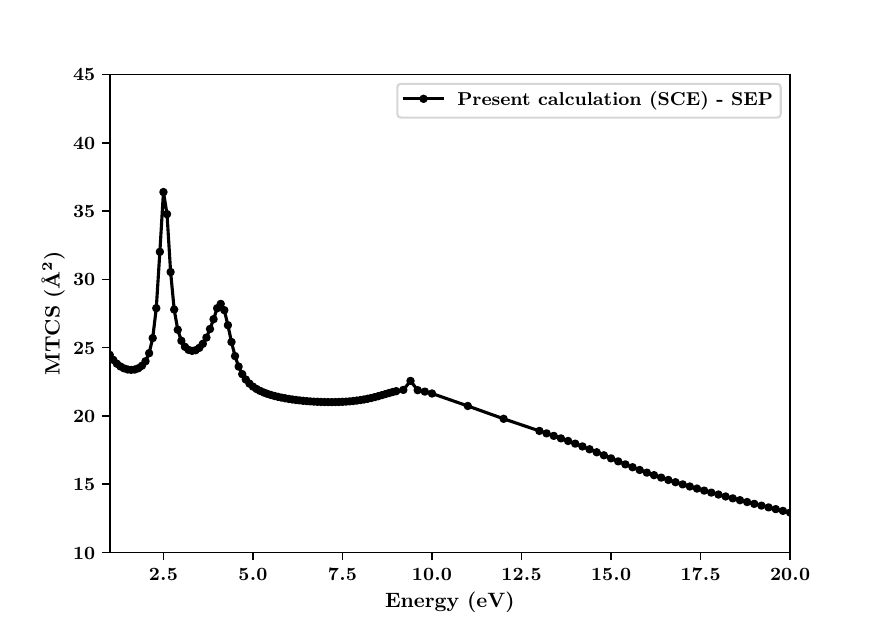}
\caption{\label{fig:10}Born-corrected momentum transfer cross-section for the scattering of electrons by 4H-pyran.}
\end{figure}

\begin{figure*} 
\begin{subfigure}{0.45\textwidth}
\includegraphics[width=\textwidth]{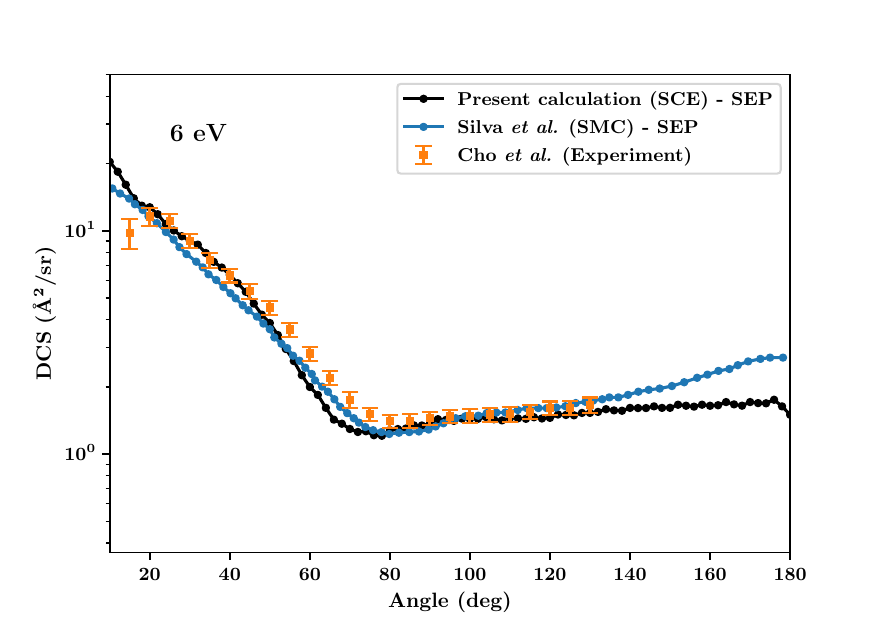}
\caption{}
\end{subfigure}
\begin{subfigure}{0.45\textwidth}
\includegraphics[width=\textwidth]{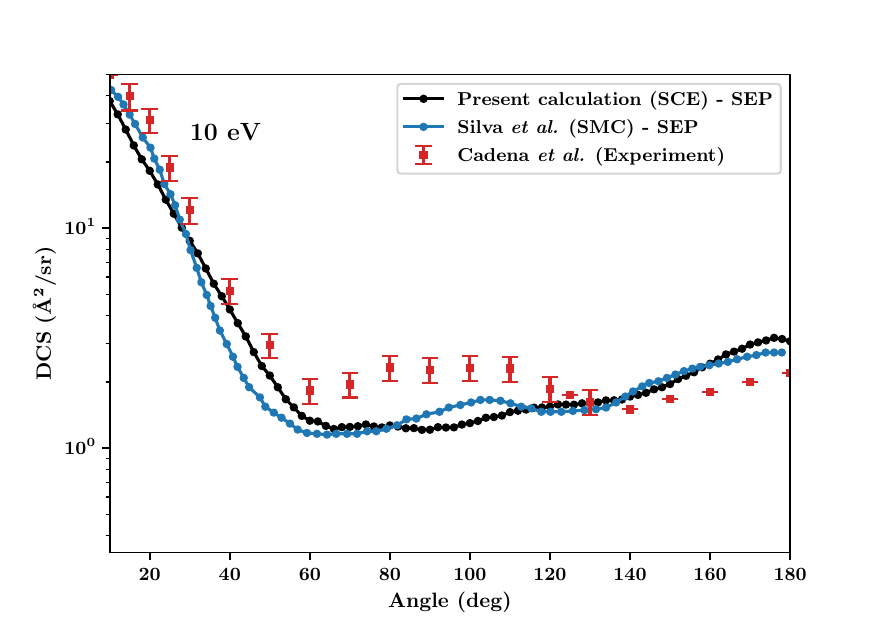}
\caption{}
\end{subfigure}
\begin{subfigure}{0.45\textwidth}
\includegraphics[width=\textwidth]{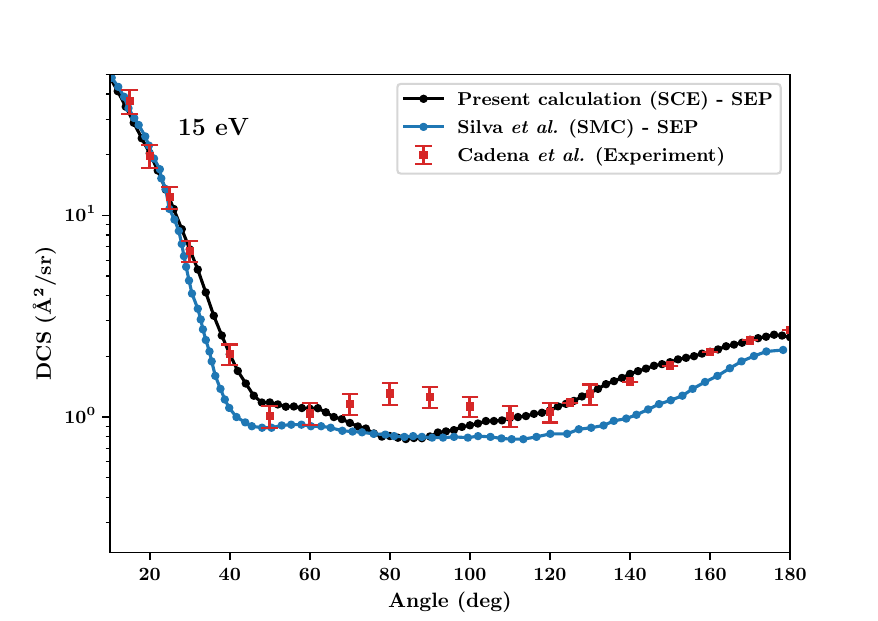}
\caption{}
\end{subfigure}
\caption{\label{fig:11} Born-corrected differential cross-section for the scattering of electrons by 4H-pyran at the incident energies of (a) 6 eV (b) 10 eV (c) 15 eV. Black dotted-line: present work. Blue dotted-line: SMC calculation of Silva \textit{et al.} \cite{silva2024elastic} under 41-channel SEP approximation for 6 eV and 10 eV DCS, and 158-channel SEP approximation for 15 eV DCS.  Red squares with error: experimental DCS measured by Cadena \textit{et al.}\cite{cadena2022cross} for benzene molecule. Orange squares with error: experimental DCS measured by Cho \textit{et al.}\cite{Cho_2001} for benzene molecule. }
\end{figure*}

\section{RESULTS AND DISCUSSION}
This section highlights the key findings of our calculations performed for the scattering of low-energy (1--20 eV) electrons from 2H-pyran and 4H-pyran molecules.
Fig. \ref{fig:2} illustrates the contribution of each symmetry component, $\rm{A^\prime}$ and $\rm{A^{\prime\prime}}$,  of the $\rm{C_s}$ point group of 2H-pyran molecule to the integral elastic cross-section (ICS). The $\rm{A^{\prime\prime}}$ irrep ICS exhibits two low-lying resonance peaks which are further analyzed utilizing the time delay approach. The maximum eigenvalue of the time delay matrix (refer to Eq. \ref{eqn15}), corresponding to the maximum time delay, at each incident energy is plotted and shown in Fig. \ref{fig:3}. The resonances manifest themselves as Lorentzian peaks in the time delay plots. For the $\rm{A^{\prime\prime}}$ irrep, the two resonance peaks identified in Fig. \ref{fig:2} are clearly visible in Fig. \ref{fig:3} of the time delay. The resonance peaks are fitted using the Lorentzian function \cite{loupas2018shape,stibbe1996time} and the width and position of each peak are displayed in Table \ref{tab:table2}. The resonances corresponding to the $\rm{A^{\prime\prime}}$ symmetry component are positioned at 1.78 eV and 5.62 eV, each having a width of 0.02 eV. We have compared our results with the ones reported by Silva \textit{et al.} \cite{silva2024elastic} using Schwinger multichannel method. Their calculations performed under 1-channel (only elastic channel is open) static-exchange (SE) approximation give the first resonance peak at 2.80 eV and the second one at 6.70 eV. Both these peaks lie higher than the ones reported by us. This is due to the absence of polarization potential in the SE calculations. Furthermore, we compared our results with those calculated by Silva \textit{et al.} \cite{silva2024elastic} under 1-channel static-exchange plus polarization (SEP) approximation using a minimum number of singly excited spin-adapted Slater determinants in order to
construct the configuration
state functions space together with the ground state. In their work, two polarization schemes have been used, however, we have considered the scheme mentioned above for comparison purposes which shows that the $\rm{A^{\prime\prime}}$ resonances are positioned at 1.32 eV and 4.46 eV. These lie lower to the resonance positions determined by us, the reason being the effect of the polarization potential. In the SCE method, a model polarization potential is used which might underestimate the potential value and hence shift the resonance peaks towards higher energies. Another resonance of 2H-pyran molecule belonging to the $\rm{A^{\prime}}$ symmetry is located at around 12 eV. In the time delay plot of Fig. \ref{fig:3}, this resonance is visible as a slight hump, and hence we could not fit it. The peak position of 12 eV is suggested by analyzing the $\rm{A^{\prime}}$ component of ICS in Fig. \ref{fig:2}. Our resonance peak lies slightly higher than the ones reported by Silva \textit{et al.} \cite{silva2024elastic} under SE (11.20 eV) and SEP (9.50--10 eV) approximations. This might be due to the fact that as the incident energy increases, the effect of polarization becomes less visible. All three resonances for 2H-pyran can be seen in the total elastic ICS, presented in Fig. \ref{fig:4}, as well. In the subfigure  \ref{fig:4} (a), we have compared our Born-uncorrected ICS result obtained in the body-fixed frame using VOLSCAT and the rotationally summed ICS in the space-fixed frame using POLYDCS with the one reported by Silva \textit{et al.} \cite{silva2024elastic} under 1-channel SE and 1-channel SEP approximations. Our body-fixed result showcases an overall good comparison with the SEP result at energies less than about 8 eV and with the SE result at higher energies, the reason again being the crucial role played by polarization potential at lower energies. However, the rotationally summed space-fixed ICS has an overall lower value throughout the energy range. This might be the outcome of multichannel coupling (here, rotationally excited channels), where the magnitude of the cross-section usually decreases due to the flux distribution between the elastic and inelastic channels. Since 2H-pyran is a polar molecule, we have also included Born correction in our calculations and the subfigure  \ref{fig:4} (b) compares our result of Born-corrected rotationally summed ICS with the one calculated by Silva \textit{et al.} \cite{silva2024elastic} using up to 129-channel (considering 128 open electronically excited states and the ground state) SEP approximation. The overall variation in the result may be because our calculation involves open rotationally excited states whereas Silva's have open electronically excited states.       

In addition to the three resonances discussed above for 2H-pyran, we have identified a potential fourth resonance peak belonging to the $\rm{A^{\prime}}$ symmetry from the time delay plot (refer to Fig. \ref{fig:3} ) positioned at 8.3 eV and having a width of 0.1 eV. This resonance is reflected in the $\rm{A^{\prime}}$ symmetry ICS of Fig. \ref{fig:2} with a broad peak at around 7.0--8.6 eV as well as the elastic ICS presented in Fig. \ref{fig:4}, with the Born-uncorrected body-fixed frame ICS having a more prominent broad peak at around 7.3 eV to 8.7 eV. This peak is not identified in the work of Silva \textit{et al.} \cite{silva2024elastic} probably due to the presence of numerous structures around 10 eV. The remaining two peaks in Fig. \ref{fig:3}, at energies 4 eV and 18.1 eV of $\rm{A^{\prime}}$ symmetry, might be the outcome of excitation thresholds \cite{loupas2019shape}, the first peak being closer to the second excited state, $\rm{2^3A^\prime}$, of 2H-pyran having vertical excitation energy of 4.14 eV. \cite{silva2024elastic} Since we are unable to assign any excited state to the second peak at 18.1 eV due to a lack of literature data on excitation energy, this peak might also be some result of numerical instability.

We also calculated the Born-corrected rotationally summed momentum transfer cross-section (MTCS) for 2H-pyran, which is presented in Fig. \ref{fig:5}. As per our knowledge, no theoretical or experimental MTCS data is available in the literature for the molecule. Fig. \ref{fig:6} illustrates the Born-corrected rotationally summed differential cross-section (DCS) results for 2H-pyran at the incident energies of 6 eV, 10 eV, and 15 eV. As the DCS values at forward or low angles contribute most to the ICS, our DCS in that angular region is in good agreement with the one reported by Silva \textit{et al.} \cite{silva2024elastic} for the energies of 10 eV and 15 eV. The effect of polarization at low energies is evident from the variation in our DCS result at 6 eV compared to Silva \textit{et al.} \cite{silva2024elastic} As experimental data of the DCS for 2H-pyran is unavailable, we have compared our result with the experimental DCS for benzene, a structurally similar molecule. For 6 eV, we have used the experimental DCS for benzene reported by Cho \textit{et al.} \cite{Cho_2001}, while, for 10 eV and 15 eV we have compared our result with Cadena et al.'s \cite{cadena2022cross} experimental DCS for benzene. Again, the experimental DCS fairly agrees with our DCS data at low scattering angles for 10 eV and 15 eV. The overall difference in the magnitude of our DCS and that of Silva's and Cadena's, especially at higher scattering angles, might originate from the lower point group symmetry of 2H-pyran as the SCE method gives better results for higher symmetry molecules. At 6 eV, the overall difference in our result and that of Silva's and Cho's can be attributed to both polarization and symmetry.                     

Moving onto the results obtained for 4H-pyran molecule, Fig. \ref{fig:7} represents the symmetry-wise ($\rm{A_1}$, $\rm{A_2}$, $\rm{B_1}$, and $\rm{B_2}$) contribution of 4H-pyran, belonging to the $\rm{C_{2v}}$ point group, to the total elastic ICS. Both $\rm{A_2}$ and $\rm{B_1}$ irreps in Fig. \ref{fig:7} features a prominent resonance peak. The same can be identified in the time delay plot of $\rm{A_2}$ and $\rm{B_1}$ irreps illustrated in Fig. \ref{fig:8}. Fitting the time delay resonance peak of $\rm{A_2}$ symmetry using the Lorentzian function, we obtained its position and width as 2.5 eV and 1.208 eV, respectively. Similarly, by fitting the resonance peak belonging to the $\rm{B_1}$ irrep, we obtained the peak to be positioned at 4.2 eV having a width of 0.1 eV. Comparing our result with the one reported by Silva \textit{et al.} \cite{silva2024elastic} using 1-channel SE approximation, we see that their resonances belonging to $\rm{A_2}$ and $\rm{B_1}$ symmetries are positioned at 3.70 eV and 5.60 eV, respectively. These lie higher than our peaks as was the case for 2H-pyran. Here, the reason again is the exclusion of polarization potential in SE calculations thereby overestimating the position of the resonances. Further comparing our result with the 1-channel SEP calculation of Silva \textit{et al.} \cite{silva2024elastic}, we see that their estimated peak positions for $\rm{A_2}$ and $\rm{B_1}$ resonances are at 2.20 eV and 3.94 eV, respectively, which are lower than ours. Again, in this case, the model polarization potential used in SCE approach may have underestimated the potential value and hence slightly overestimated the resonance peak positions. 

The two resonances discussed above are clearly visible in the total elastic ICS shown in Fig. \ref{fig:9}. Fig. \ref{fig:9}(a) illustrates the Born-uncorrected ICS in the body-fixed frame calculated using VOLSCAT as well as the Born-uncorrected rotationally summed ICS obtained in the space-fixed frame from POLYDCS. These two are compared with the 1-channel SE and SEP ICS results of Silva \textit{et al.} \cite{silva2024elastic} Below the energy of around 10 eV, our ICS data is overall in close agreement with the 1-channel SEP result of Silva. Above 10 eV, as the effect of polarization becomes less evident, both the SE and SEP results of Silva agree well with our data. It is again noticed that the born-uncorrected rotationally summed ICS calculated by us, using POLYDCS, as a whole has a lower amplitude than the result obtained using VOLSCAT. This can be attributed to the flux distribution between elastic and inelastic channels leading to the decrement of cross-section value. Born correction was also included in the calculations as 4H-pyran is a polar molecule. Fig. \ref{fig:9}(b) shows the comparison of the Born-corrected rotationally summed ICS calculated by us with the one calculated by Silva \textit{et al.} \cite{silva2024elastic} under up to 249-channel (considering 248 open electronically excited states and the ground state) SEP approximation. Both the results show a good agreement, however, our ICS data has a higher value almost throughout the energy range. This might be due to different open channels considered during calculations, with rotational open channels in our case and electronic open channels in Silva's case.      

Another resonance belonging to the $\rm{B_2}$ symmetry can be identified as a peak in the symmetry-wise ICS contribution of Fig. \ref{fig:7} at around 14.2 eV. In the time delay plot of Fig. \ref{fig:8} the resonance peak is at 14.4 eV. The width of the peak was determined to be 0.2 eV. 
This resonance can be seen as a broad peak, around 12.5--16 eV, in the total ICS result presented in Fig. \ref{fig:9}. This is a bit higher than the resonance position reported by Silva \textit{et al.} \cite{silva2024elastic}, at 10.60 eV under 1-channel SE and 9.50 eV to 10 eV under 1-channel SEP approximations. A clear comparison of the resonance positions and widths of all the resonances calculated by us with the resonance positions reported by Silva are displayed in Table \ref{tab:table3}. 

Furthermore, we can locate another broad resonance around 8.5 eV to 10 eV in Fig. \ref{fig:9}. To assign a symmetry to this resonance, we analyzed the irrep-wise ICS of Fig. \ref{fig:7} and the time delay plot of Fig. \ref{fig:8}. Fig. \ref{fig:7} shows a peak around 8.9 eV of $\rm{B_2}$ symmetry and a broad peak around 9 eV to 13 eV belonging to the $\rm{A_1}$ symmetry. However, the corresponding peak belonging to $\rm{B_2}$ symmetry is not visible in the time delay plot of Fig. \ref{fig:8} (hence unphysical), we could only identify the peak corresponding to the $\rm{A_1}$ symmetry at 9.4 eV with a width of 0.2 eV. As discussed earlier, the time delay method gives a more valid result for resonances, we correlate the broad resonance peak visible in Fig. \ref{fig:9} to $\rm{A_1}$ symmetry. No resonance peak belonging to the $\rm{A_1}$ symmetry has been reported by Silva \textit{et al.} \cite{silva2024elastic} The remaining peaks in Fig. \ref{fig:8} at 3.9 eV and 18.2 eV belonging to $\rm{A_1}$ and $\rm{B_2}$ symmetry, respectively might be the outcome of excitation thresholds.\cite{loupas2019shape} The 3.9 eV peak lies close to the second excited state $\rm{1^3A_1}$ of 4H-pyran having a vertical excitation energy of 3.58 eV. \cite{silva2024elastic} The 18.2 eV peak is similar to the one obtained for 2H-pyran and therefore might also be the result of some numerical instability.

Fig. \ref{fig:10} represents the Born-corrected rotationally summed momentum transfer cross-section (MTCS) of the 4H-pyran molecule. There is no data, theoretical or experimental, available in the literature to provide a comparison. We have also presented the Born-corrected rotationally summed differential cross-section (DCS) for three different incident energies of 6 eV, 10 eV, and 15 eV in Fig. \ref{fig:11}. Our DCS is in a better comparison with the one reported by Silva \textit{et al.} \cite{silva2024elastic}, for all three energies, owing to the higher point group symmetry of 4H-pyran in comparison to 2H-pyran. As done for 2H-pyran, we have compared our DCS for 4H-pyran with the experimental DCS reported by Cho \textit{et al.} \cite{Cho_2001} for benzene at 6 eV and that reported by Cadena \textit{et al.} \cite{cadena2022cross} for benzene at 10 eV and 15 eV. The experimental data also compares well with our DCS for all the energies, with our data showing a better agreement with the experiment as compared to Silva's at 15 eV.     

\begin{figure*} 
\begin{subfigure}{0.45\textwidth}
\includegraphics[width=\textwidth]{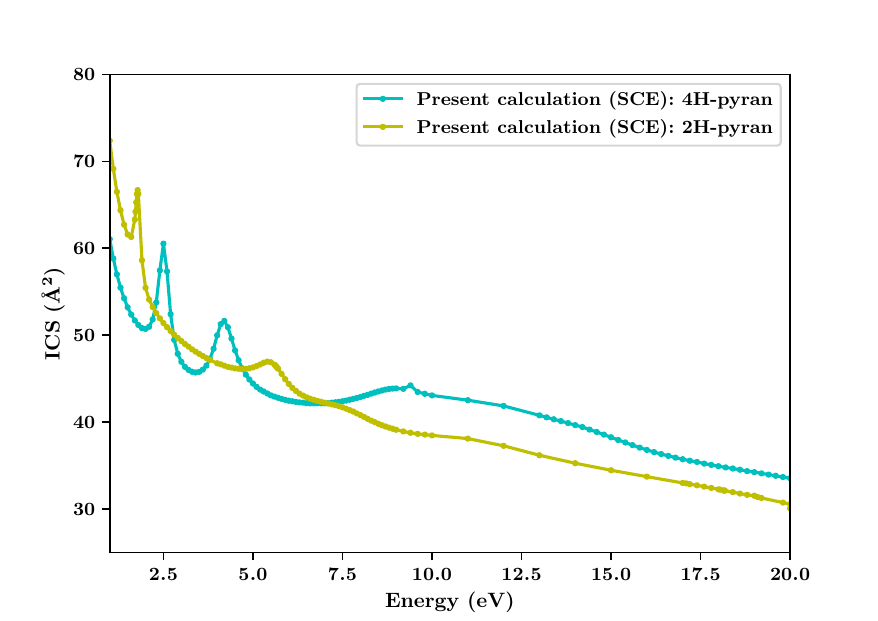}
\caption{}
\end{subfigure}
\begin{subfigure}{0.45\textwidth}
\includegraphics[width=\textwidth]{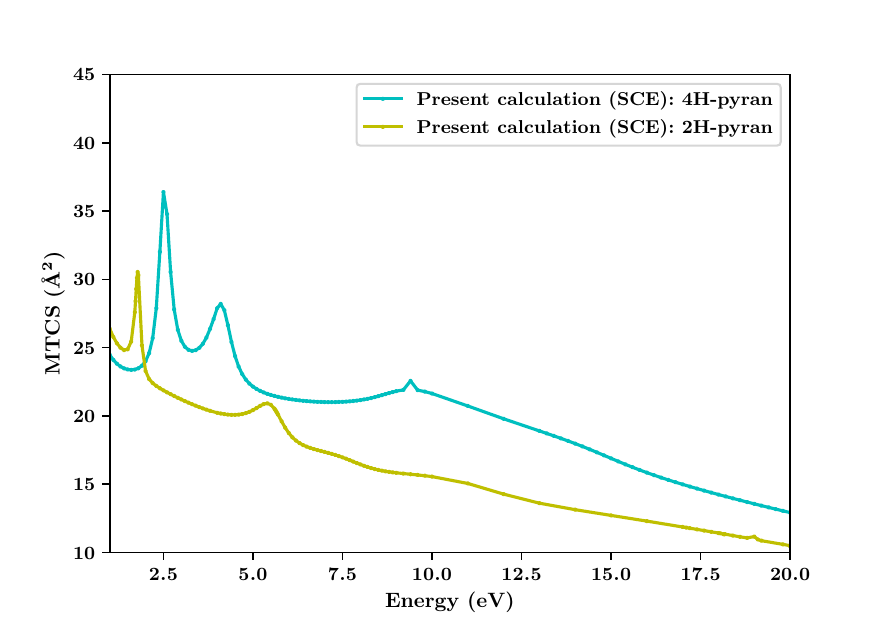}
\caption{}
\end{subfigure}
\begin{subfigure}{0.45\textwidth}
\includegraphics[width=\textwidth]{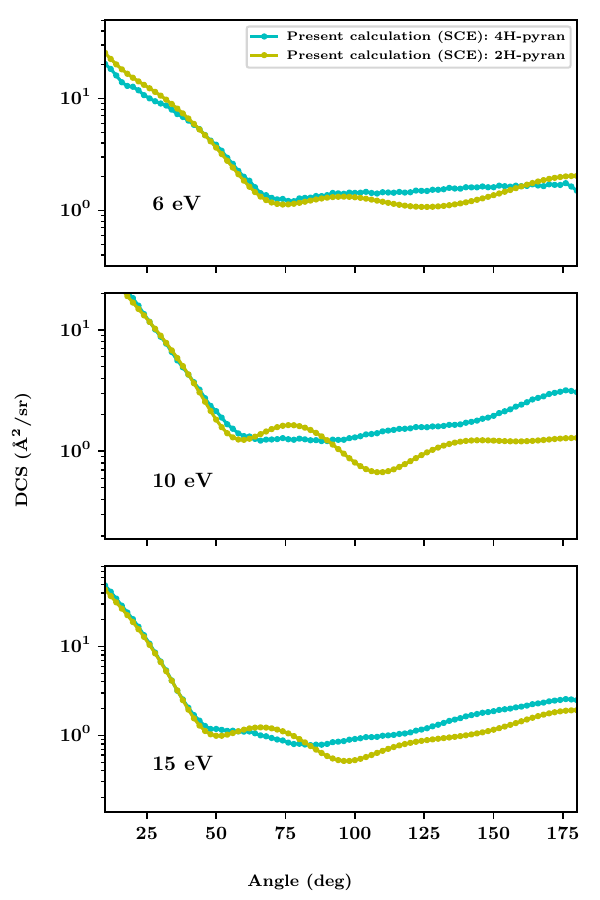}
\caption{}
\end{subfigure}
\caption{\label{fig:12} Comparison of the Born-corrected (a) integral cross-section, (b) momentum transfer cross-section, and (c) differential cross-section of 2H-pyran (yellow dotted-line) and 4H-pyran (cyan dotted-line). }
\end{figure*}

Comparing the resonance peak positions listed in Table \ref{tab:table2} and Table \ref{tab:table3} for the isomers 2H-pyran and 4H-pyran, respectively, we could verify the isomeric effect leading to a difference in positions of the resonance peaks for the two molecules, as highlighted in Fig. \ref{fig:12} and reported by Silva \textit{et al.} \cite{silva2024elastic} in their work. In Fig. \ref{fig:12}, the difference in the magnitudes of the ICS, MTCS, and DCS for the two isomers is the combined effect of the difference in positioning of the double bonds, the different symmetries, as well as slightly different dipole moments of 2H-pyran and 4H-pyran.

\section{CONCLUSIONS}
In this work, we reported the integral (ICS), differential (DCS), and momentum transfer (MTCS) cross-sections for the elastic scattering of low-energy electrons from two isomers, 2H-pyran and 4H-pyran. Moreover, we presented the analysis of the resonance peaks using the time delay approach. The single-center expansion method was employed to carry out the calculations. Analyzing the time delay graph, which plots the maximum eigenvalue of the time delay matrix versus the incident energy, we were able to identify three resonance peaks for each 2H-pyran and 4H-pyran molecule. The positions of these peaks were in good accord with those reported by Silva \textit{et al.} \cite{silva2024elastic} under SE and SEP approximations of the Schwinger multichannel approach, particularly the two low-lying resonances. We identified a fourth potent resonance peak, for both 2H-pyran and 4H-pyran, positioned at 8.3 eV ($\rm{A^{\prime}}$ symmetry) and 9.4 eV ($\rm{A_1}$ symmetry), respectively, not reported by Silva \textit{et al.} \cite{silva2024elastic} in their work. Future investigations on these molecules, both on the theoretical and experimental front, are suggested to testify to our result. 
Our calculated DCS data, which essentially serves as a stringent test to validate the applicability of any method, were in good agreement with that of Silva's and the experimental data of Cho's and Cadena's, especially for the 4H-pyran molecule which exhibits a higher level of symmetry. We also reported for the first time, the MTCS data for both 2H-pyran and 4H-pyran, which is utilized in studying the electron transport properties in gases.       

\begin{table}
\begin{threeparttable}
\caption{\label{tab:table2} Width and position of the resonances for 2H-pyran (Units in eV) obtained using time delay analysis. \cite{loupas2018shape,stibbe1996time}  }
\begin{ruledtabular}
\begin{tabular}{ p{1.8cm} p{1.5cm} p{1.5cm} p{1.5cm}}
Symmetry component& Resonance width & Resonance position   & Resonance position, Silva\textit{ et al.} \cite{silva2024elastic}\\
\hline
$\rm{A^{\prime\prime}}$ & 0.02 & 1.78  & 2.80 (SE), 1.32 (SEP) \\
$\rm{A^{\prime\prime}}$ & 0.02 & 5.62  & 6.70 (SE), 4.46 (SEP)\\
$\rm{A^\prime}$&- & 12\tnote{$^*$} & 11.20 (SE), 9.50-10.00 (SEP)\\
$\rm{A^\prime}$& 0.1 & 8.3 & -\\
\end{tabular}
\begin{tablenotes}
    \small
    \item [$^*$]obtained by analyzing the ICS of Fig. \ref{fig:2}  
    
\end{tablenotes}
\end{ruledtabular}
\end{threeparttable}
\end{table}

\begin{table}
\caption{\label{tab:table3} Width and position of the resonances for 4H-pyran (Units in eV) obtained using time delay analysis. \cite{loupas2018shape,stibbe1996time}  }
\begin{ruledtabular}
\begin{tabular}{ p{1.8cm} p{1.5cm} p{1.5cm} p{1.5cm}}
Symmetry component& Resonance width & Resonance position   & Resonance position, Silva\textit{ et al.} \cite{silva2024elastic}\\
\hline
$\rm{A_2}$ & 1.208 & 2.5  & 3.70 (SE), 2.20 (SEP) \\
$\rm{B_1}$ & 0.1 & 4.2  & 5.60 (SE), 3.94 (SEP)\\
$\rm{B_2}$& 0.2 & 14.4 & 10.60 (SE), 9.50-10.00 (SEP)\\
$\rm{A_1}$& 0.2 & 9.4 & -\\
\end{tabular}
\end{ruledtabular}
\end{table}

\begin{acknowledgments}
D.G acknowledges the Science and Engineering Research Board (SERB), Department of Science and Technology (DST), Government of India (Grant No. SRG/2022/000394) for providing financial support and computational facility.
\end{acknowledgments}

\section*{Data Availability Statement}

\textcolor{black}{The data that support the findings of this study are available upon reasonable request from the authors.}

\section*{Author Contributions}
\textbf{Snigdha Sharma:} Conceptualization (lead), Data curation (lead), Investigation (lead),  Methodology (lead), Resources (Supporting), Validation (lead), Writing – original draft (lead), Writing – review \& editing (equal); \textbf{Dhanoj Gupta:} Conceptualization (equal), Data curation (equal), Investigation (equal),  Methodology (equal), Resources (lead), Validation (equal), Writing – original draft (equal), Writing – review \& editing (lead), Supervision (lead).

\clearpage
\section*{REFERENCES}
\nocite{*}
\bibliography{aipsamp}

\end{document}